\documentclass[12pt,preprint]{aastex}

\shorttitle{FBS 0107-082}
\shortauthors{Honeycutt \& Kafka}

\begin{document}


\title
{FBS 0107-082: A Symbiotic Binary in a Rare Prolonged 
Outburst?}





\author{R.K. Honeycutt}
\affil{Astronomy Department, Indiana University, Swain Hall West, 
Bloomington, IN 47405 USA}
\and
\author{S. Kafka}
\affil{Dept.of Terrestrial Magnetism, Carnegie Inst. of Washington,
5241 Broad Branch Road NW, Washington, DC 20015 USA}

\begin{abstract}

FBS0107-082 is an emission line object previously classified as 
a nova-like cataclysmic variable star.  New optical spectroscopy
shows very strong hydrogen Balmer lines, along with a nebular
forbidden line spectrum and absorption features from an early-F
photosphere.  When combined with other IR and optical data from 
the literature, these data point to the object being a symbiotic nova 
seen in a prolonged outburst.  Photometry on time scales of minutes,
days, and years show only very weak variability.  

\end{abstract}

\keywords{Stars-binaries:symbiotic-stars:individual(FBS0107-082)}

\section{Introduction}

FBS~0107-082 (also known as Cet 7, and designated FBS0107 in this paper)
was found in the First Byurakan Survey (FBS)
for objects with UV excesses.  Based on spectra obtained with the 6-m
telescope at the Special Astronomical Observatory, Kopylov et al. (1988)
provided a preliminary classification as a nova-like (NL) cataclysmic
variable (CV).  FBS0107 was subsequently listed as a CV in SIMBAD and in
the Downes, Webbink, \& Shara (1997) CV catalog, but has received very little
attention since the initial discovery paper.  We obtained spectra of FBS0107 
in 2005 as part of a survey for winds in NL CVs, finding that the spectrum 
resembles that of a symbiotic star or planetary nebulae.  We have complemented
these initial spectra with multi-epoch photometry and with an analysis of
archival ESO spectra.

In this paper we report on our spectroscopic and photometric findings and
discuss the most likely kinds of objects that fit these findings.
The star appears to be an unusual symbiotic binary.  Symbiotic stars 
(Kenyon 1986) have evidence for
a hot component in the near UV plus a cool component in the near-IR.
In most cases the hot component is judged to be a white dwarf or
evolved subdwarf, and the
companion to be an M giant with a strong wind.  Nebular emission lines are 
produced by the action of the radiation field of the hot component on the 
surrounding plasma from the wind.  Symbiotic stars are not a particularly
homogeneous group and the evolutionary state of the hot component (e.g.
subdwarf or white dwarf) is often uncertain.  Our observations are not
a perfect fit to the symbiotic classification, but this identification
seems to best match the properties of FBS0107.    
 
\section{Data Acquisition and Reduction}

\subsection{Photometry}

V-band photometry was obtained over a 1.5 month interval in 2005 on the 0.8-m 
telescope of Tenagra Observatory\footnote{http://www.tenagraobservatories.com/}.  
There were typically 1-4 exposures on most clear nights,
but on two nights we obtained nearly continuous coverage for several
hours.  These data were reduced using a custom pipeline consisting of 
IRAF\footnote{IRAF is distributed by the National Optical Astronomy
  Observatories, which are operated by the Association of Universities for
  Research in Astronomy, Inc., under cooperative agreement with the National
  Science Foundation.} 
routines for detector calibrations, followed by the application of 
SExtractor\footnote{SEextractor is a source detection and photomery package
  described by Bertin and Arnouts 1996.  It is available from 
  http://terapix.iap.fr/soft/sextractor/.}.  The light curves were then generated 
using the incomplete ensemble photometry technique contained in 
Astrovar\footnote{ASTROVAR is a custom Indiana
  University package based on the technique described in Honeycutt 1992, but 
  with the addition of a graphical user interface.}.  
FBS0107 was also placed on the
long-term monitoring program of the Indiana University (IU) 1.25-m autonomous 
  telescope\footnote{This facility is located 20 km north of Bloomington IN,
  in the Morgan-Monroe State Forest}
for $\sim$ 16 months in 2007/2008.  The useable data are comprised of 122 
exposures on $\sim$113 different nights.  The 1.25-m data were also reduced using 
the IRAF/SExtractor/Astrovar pipeline.  Table 1 is a log of our photometric
data.  Column 1 assigns a number to identify each data set, column 2 gives 
the UT dates, column 3 the telescope used, column 4 the exposure time in sec, 
column 5 the number of useable exposures, column
6 the duration of the sequence, and column 7 the typical spacing of the
exposures.

Both the Tenagra exposures and the IU 1.25-m exposures were compromised
by various circumstances, and did not reach the desired (and expected) level of
precision.  The Tenagra calibration data is acquired automatically, with
no user control of the process.  In our case the twilight flats, acquired
on most evenings, were very underexposed; we ended up using a median
combination of all the twilight flats over the $\sim$1.5 months of the
FBS0107 Tenagra exposures.  Furthermore the Tenagra bias levels were unstable,
sometimes leading to negative sky counts. The IU 1.25-m data and calibrations were
also acquired in unattended fashion, during commissioning of the telescope and
detector systems.  Few 
twilight flats were acquired, and the nightly dome flats gave poor results.
Therefore we used median sky flats constructed from all the exposures on a
given (or adjacent) night.  Also, we had significant dark current from the
thermoelectically-cooled system that was in service on the IU 1.25-m at this time, 
which further degraded the S/N.  Fortunately Astrovar has sufficient tools to 
evaluate errors 
that might result from these difficulties (such as non-linearity, or field errors),
and Astrovar also robustly estimates external errors based on the repeatability
of all constant stars in the field.  

In the end the average error for stars with
brightness similar to FBS0107 was $\sim$0.02 mag for the Tenagra data
and $\sim$0.025 mag for the IU 1.25-m data.  These errors are for differential
magnitudes with respect to the ensemble.  There are no secondary standards
available for this field (although the AAVSO is in the process of establishing
such standards) so we used USNO-B1 magnitudes\footnote{The USNOFS Image and 
  Catalogue Archive is operated by the United States Naval Observatory, Flagstaff 
  Station (http://www.nofs.navy.mil/data/fchpix/).}
to establish a somewhat crude, but nevertheless realistic, zeropoint.  
The average of the B2 and R2 
USNO-B1.0 magnitudes for several stars near FBS0107 were compared to the Astrovar 
differential V magnitudes of those same stars, separately for the Tenagra and
for the IU 1.25-m data, to find the
correction from Astrovar magnitude to V.  This process has a formal
uncertainty of 0.06 (sdm) mag, but the zeropoint error could be as large as 0.15 mag
considering possible systematic errors in USNO-B1.0, plus errors introduced by
our averaging of B and R to estimate the V magnitude.  

\subsection{Spectroscopy}

Five spectra over $\sim$1 hr were obtained in 2005 using MOS/Hydra multiple 
object spectrograph on the 
WIYN\footnote{The WIYN Observatory is a joint facility of the University of 
  Wisconsin-Madison, Indiana University, Yale University, and the National 
  Optical Astronomy Observatory} 
telescope.  Grating 600 was used in first order, providing
coverage $\sim$5300-8200\AA.  The ``red'' 2$\arcsec$ fiber bundle was employed,
yielding $\sim$3\AA\ resolution; numerous other fibers were used for sky
subtraction.  Also, a single spectrum
was obtained in 2006 using the RC slit spectrograph on the Kitt Peak  
4-m telescope\footnote{The Mayall 4-m telescope is at
  Kitt Peak National Observatory, a division of the National Optical Astronomy
  Observatory, which is operated by the Association of Universities for Research
  in Astronomy, Inc., under cooperative agreement with the National Science
  Foundation}.
Grating KPC-007 was used in first order, providing coverage $\sim$5600-7900\AA\ 
at $\sim$3\AA\  resolution.  Finally, the archives of the ESO 
Observatory\footnote{
  Based on observations made with the European Southern Observatory
  telescopes obtained from the ESO/ST-ECF Science Archive Facility.}
were found to contain spectra of FBS017 acquired during a single night in 2004
using the EFOSC-2 spectrograph on the 3.6-m telescope at La Silla.  Eighteen 
spectra were available using grism \#18, covering $\sim$4700-6700\AA\, along
with 5 spectra using grism \#7, with useable coverage 3700-5240\AA.  A 1$\arcsec$ 
slit was used, and the instrument manual gives a resolution of 
7.4\AA\ for these settups.  However, we find that the narrowest emission lines
in the FBS0107 spectra have FWHM $\sim$5-6\AA.
For detector calibrations we used standard IRAF
procedures, and for spectral extractions and wavelength calibrations we used 
IRAF's onedspec/twodspec packages.  No spectrophotometric calibrations were
applied for any of the spectra, and the continua in the reduced WIYN and
Mayall  spectra were nomalized to unity.  Also, no corrections were made for 
telluric spectral features in the near-IR.  Table 1 is a log of the 
spectrographic observations.

\section{Analysis}

\subsection{Photometry}

Figure 1 plots the results from the two nights of nearly continuous
monitoring (Sets 1 and 2 in Table 1).  The top plot appears to contain a small 
outburst of amplitude
$\sim$0.07 mag, while the bottom plot shows a decline of $\sim$0.04 mag 
over 5 hours.  The Astrovar light curves of several field stars of similar
brightness to FBS0107 were examined and did not show such
features.  The features are therefore probably real, but since they
barely exceed the errors in the light curves we are reluctant to draw
any conclusions from them.  The variability over a night is
smaller than expected from CV-like flickering from accretion.

Figure 2 shows the longer-term light curve of FBS0107.  The top panel contains 
all the data in Sets 1, 2, and 3 of Table 1, while  the middle and lower 
panels contain the data for the two observing seasons in Set 4.
It seems clear from Figure 2 that there is no systematic long-term trend over 
these 4 years.  Nevertheless the scatter generally exceeds the individual 
error bars by 3-4$\times$, at least for the points having smaller errors.
To further illustrate this effect, Figure 3 is like Figure 2 except that 
data points having errors $>$0.035 mag are excluded (these were obtained
under partly cloudy conditions).  Also, the
two sequences over single nights have been averaged to one data point each
for Figure 3.  One can find in Figure~3 a number of instances of systematic drifts 
in both directions amounting to $\sim$0.1 mag over 15-20 days.  

There is an indication of much longer time-scale variability from the
magnitudes in USNO-B1.0.  The average of the B and R magnitudes in USNO-B1.0
for FBS0107 is 14.72 (epoch 1951-1954).  This is 0.33 mag brighter than our 
average magnitude of 15.05 in 2005-2008.  One might question this result 
because of the
uncertain effect of averaging B and R to estimate V.  However, the
exercise is robust in the following sense:  The magnitude differences between
our Astrovar instrumental magnitudes and the B,R average for our secondary
standards are consistent to within 0.1 mag (s.d.s.o.), whereas the similar
magnitude difference of FBS0107 is off by 0.33 mag from the mean of the
other field stars.  In other words, if one tries to treat FBS0107 as a secondary 
standard, it deviates from the field stars by more than 3-sigma, probably because 
it is variable.  FBS0107 has a B-R color that is similar to that of the
secondary standards, so we conclude that FSB is now fainter by $\sim$0.3 mag in
V compared to when the Palomar plates were exposed $\sim$55 years earlier.
 
To summarize the photometry, we find that FBS0107 is a variable, but a
puny one.   We see changes of a 
few hundredths of a mag during a night and perhaps 2-3$\times$ that amount 
over a few weeks, but with no systematic trends exceeding 0.05 mag over our
4 year interval of photometry.
These variations do not exceed our observational errors by much, and
we are therefore unable to characterize the changes very well.  No flickering
was detected, nor any indication of an eclipse.

\subsection{Spectroscopy}

No exposure-to-exposure changes were apparent in the line strengths of the
reduced spectra from any given night.  Furthermore we did not detect any 
radial velocity variations during these sequences, to within
$\sim\pm$10 km s$^{-1}$ for 5 WIYN spectra, and to within
$\sim\pm$15 km s$^{-1}$ for the ESO spectral sequences.
Therefore we median-combined the 5 WIYN spectra, the 18 ESO red spectra, 
and the 5 ESO blue spectra, for further analysis.

First we compare the WIYN and KPNO 4-m spectra.  These are separated by
2 months but have similar resolutions and wavelength ranges.
Figure~4 shows the average of the 5 WIYN spectra of FBS0107 along with an
inset plot of the H$\alpha$ line on an expanded scale, while Figure~5 is a 
similar plot for the KPNO 4-m spectrum.  In each case H$\alpha$ dominates
the spectrum, seen at a rather spectacular $\gtrsim$50$\times$ the continuum 
level, at this resolution.  The  H$\alpha$ line is steeper on the red side, 
with this
asymmetry being more pronounced in the KPNO 4-m spectrum.  The weaker
lines are shown in Figures~6 and 7, for two adjacent spectral regions in
common to the WIYN and the 4-m spectra.  Except for higher S/N in the
WIYN spectra (due to having combined 5 exposures), the pairs of spectra in 
Figures~6 and 7 (separated by 2 months) appear identical.  Most of the lines 
are due to HeI and OI.  There is one absorption feature
near 7772\AA\, which we assign to OI. 

Figure~8 shows the median-combined ESO spectra (both red and blue).  The
identifications of the strongest lines are given in Table 3, where the
equivalent width measures (EW) have a typical error of $\sim\pm$0.5\AA.
The lines are mostly emission lines of Hydrogen and HeI, plus
forbidden emission lines of oxygen and iron.  We also see weak absorption
features at Ca H/K and at NaD.  The inset shows the Balmer discontinuity
in strong absorption.  In order to preserve information about the Balmer
discontinuity, the continua of the ESO spectra have not been normalized
to unity as was done for the WIYN and KPNO spectra.  A broad emission feature
near 6160-6170\AA in Figs. 6 and 8 is assigned to FeI 6164, likely blended
with an unidentified contributor which increases the breadth.

\section{Discussion}

The 1984 spectrum of FBS0107 in Kopylov et al. (1988) covers the range
3240-5040\AA\  at a resolution of 4.5\AA.  It shows narrow Balmer emission
lines with a steep decrement.  HeII 4686 is weak, but [OIII] 5959, 5007,
and [FeII] are strong.  The Balmer discontinuity is not apparent in the
Kopylov et al. spectrum. Their detector response is apparently weak
below 3650\AA\ but a Balmer discontinuity in emission can be ruled out.
Overall, the 1984 spectrum appears very similar
to the ESO 2004 spectrum, indicating little variability in the spectrum of the
star over those 20 years.

We discuss below a number of kinds of emission line objects as 
candidates for the nature of FBS0107, noting the similarities and
and differences to the currently-known properties of FBS0107.  We begin
with symbiotic stars. 

\subsection{Symbiotic systems}

Symbiotic stars (Friedjung \& Viotti 1982; Kenyon 1986; Allen 1988; 
Mikolajewska et al. 1988) 
are interacting binaries with orbital periods of hundreds of days, consisting of
a late-type giant doner star transfering gas (generally via a wind) to a much
hotter compact companion.  The hot object appears to most often be a
white dwarf or hot subdwarf, but may be a main sequence star or neutron star in some
instances.  The hot star photoionizes the wind, giving rise
to a nebular spectrum with emission lines due to species such as
HI, He I, He II, [OIII], [Fe II], etc.  Symbiotic stars are a rather hetrogeneous
group, and might better be considered as a common phenomenon rather
than a particular configuration and/or evolutionary state.  The FBS0107 emission 
lines  seen in Kopylov et al. (1988) and in Figures~4-8 of this paper
are generally consistent with those of a symbiotic star, albeit having somewhat
lower ionization level than average.  The Balmer decrement in FBS0107
is quite steep, as is often found in symbiotic stars.  However, because
our spectra are not on a flux scale we cannot reliably quantify the decrement. 

Apart from the emission lines, the spectrum of FBS0107 differs significantly 
from that of a classical
symbiotic star.  In the near-IR, the spectra of symbiotic stars show features
from the late type companion.  This is typically an M giant, but can be
as early as G5 (Belczy\'{n}ski et al. 2000); no such features are seen in
FBS0107.  The few absorption features we see in FBS0107 do not belong
to a late-type star. Using the plots and tables in Jaschek \& Jaschek (1995)
for the variations of the EWs of absorption features with spectral type,
we find the following constraints on the absorption line spectral class for 
FBS0107:  EW = 0.9\AA\ for CaII 3934 is most consistent with A0-A5; 
EW = $\sim$0.5\AA\ for NaD is most consistent with
A5-G5, and EW = 1.2\AA\ for OI 7772\AA\ is most consistent with
A0-F0 III-I. (This OI line is strongly luminosity sensitive,
reaching EW$\sim$2\AA\ in supergiants (Parsons 1964)).  
Overall the results show that a late A to early F photospheric absorption
spectrum is present in FBS0107.
Also, the Balmer continuum is in absorption 
in FBS0107.  In nearly all other cases in which the Balmer discontinuity is 
visible in a symbiotic, the jump is part of the nebular spectrum and appears 
in emission (see the compilations of spectra of symbiotics in Allen (1983) and
in Munari \& Zwitter (2002)).  
  
The Barbier \& Chalonge system of spectrophotometric
classification uses the parameter D to measure strength of the Balmer jump, 
where D is the log of the ratio of the fluxes on the red and blue sides of 
the discontinuity.  From the spectrum in Figure~9 we measure D = 0.30.  A 
convenient calibration of D with spectral
type can be found in Golay (1974).  For giants, D = 0.3 corresponds to
either B5 or F1, and F1 is consistent with our earlier deduction of a late A to
early F spectrum in FBS0107.

We also note that two emission features at $\lambda\lambda$6825 
and $\lambda\lambda$7082 are not present in our spectra.  These two features 
are probably due to Raman scattering of OIII resonance lines by neutral
hydrogen (Schmid 1989) and appear in most high excitation symbiotics.
This distinction may not be compelling however, because FBS0107 has a
quite low excitation emisson line spectrum.  

IR photometry can potentially be used to discriminate between symbiotic binaries 
and other sources such as planetary nebulae and CVs, as well as between
S-type and D-type symbiotics.  Using 2MASS photometry, Figure~9 is a plot of 
J-H vs H-K$_{S}$ for CVs (Hoard et al. 2002), symbiotics (Phillips 2007),
and PN (Ramos-Larios \& Phillips 2005).   
We see that the location of FBS0107 falls firmly among the
PN, not the symbiotics, mostly because the J-H color is bluer than for
most symbiotics.  Most symbiotics contain an M giant, which contributes
to making the J-H color redder.  If FSB0107 is a symbiotic, then either 
the M star is heavily dust-obscured in the optical, or the M star is
otherwise hidden.
 
Proga, Mikolajewska \& Kenyon (1994) found that the ratio of the strengths
of the He~I emission lines $\lambda$6678/$\lambda$5876 fall in the
range 0.15-0.3 for D-type (dusty) symbiotics, and 0.3-2.0 for S-type, which is 
attributed to a systematic difference in densities for the line forming 
regions in the two types.  This ratio is $\sim$0.4 in FBS0107 (see Table 3).  
This value is an upper  limit because there has been no correction for 
differential extinction between  the wavelengths of the two lines.  If 
FBS0107 is a symbiotic star, it appears 
that this ratio is consistent with either D or S-type. 

The variability properties of FBS0107 seem consistent with FBS0107
being a symbiotic star, but offer no particular evidence in favor.  The lack
of detectable radial velocity variations is not surprising given the
long orbital periods and relatively small amplitudes of the radial velocities
in symbiotic binaries.  The variability in the H$\alpha$ profile seen 
in FBS0107 is also common in symbiotic binaries, but not uniformly present.  
Likewise flickering (not seen in FBS0107) is common (Belczy\'{n}ski et al. 2000) 
but has not been reported for all symbiotics. 

Finally, symbiotic stars are often associated with a resolved nebula. 
The raw long-slit KPNO 4-m spectra were carefully examined for any
extension of the emission lines along the slit, with negative results. 
The forbidden lines in FBS0107 indicate that a nebular emission line emission
region is present in FBS0107, but the angular size of the nebula must be
quite small.

The emission line species seen in FBS0107 require
photons with energies $\sim$35 ev for ionization, which in turn
requires temperatures of $\gtrsim$10$^{5}$ $\arcdeg$K.  The early F star
seen in the spectrum
cannot be the source of this ionizing radiation, instead requiring the presence of
a white dwarf or hot subdwarf.   In principle the F star could be the
mass doner and supplier of the nebular gas (via a wind), but this would 
require considerable extension of the definitions and  mechanisms of symbiotic 
binaries. 

A substantial fraction of symbiotic binaries undergo outbursts lasting weeks to
decades, with a brightness increase of 2-7 mag.  These are thought to be due
to thermonuclear burning of hydrogen in the white dwarf envelope, analogous to
the outbursts of ordinary novae, but much slower.  Like ordinary novae, these 
symbiotic novae often display a A-F  absorption spectrum during the early outburst stages, 
due to an optically thick envelope that
develops around the erupting star.  AG~Peg, RT~Ser, and RR~Tel (see discussions and
refs in Kenyon 1986) are examples of symbiotic novae with A-F absorption line spectra
at maximum.  The outbursts can sometimes last for decades, and for more than a century 
in the case of AG~Peg.  In each of these three systems, near outburst maximum, the M star 
spectrum is not seen in the near-IR.  Emission lines appear during
the decline from maximum, with the emission lines evolving to higher excitation as the
decline progresses.  (The AG~Peg emission lines also showed P~Cygni profiles
and multiple velocity components, behaviors similar to that of ordinary novae.)  Our 
spectra of FBS0107 are mostly consistent with that of the three symbiotic nova mentioned 
above, as displayed during the early portion of the decline 
from maximum.  At this stage they have retained
some of the absorption lines from near maximum, and have developed a relatively
low excitation emission line spectrum, prior to the development of high-excitation
emission lines and the emergence of the spectrum of the M star during late decline.  
All three of the symbiotic novae mentioned above showed the Raman emission feature at 
$\lambda\lambda$6825, which is not present in FBS0107;  this could be due to FBS0107 
having not yet developed a high excitation emission line spectrum.

The photometry of FBS0107 raises two difficulties with the symbiotic nova
interpretation.  The first is the almost total lack of variability, at least by
symbiotic star standards.  All symbiotic systems (including symbiotic novae) seem to have
significant photometric variability on many time scales, with many different origins 
(e.g. Skopal 2003), while conspicious variability seems to be missing in FBS0107.
The second issue concerns the time scale of the conjectured FBS0107 outburst.
The slowest symbiotic nova, AG Peg, has a mean decline of 2.3 mag per century (Kenyon
1986).  We cannot rule out the possibility that the 1951-54 Palomar Schmidt exposures
were acquired during a separate earlier outburst.  But if FBS0107 has remained bright
since 1951-54, then 
the decline was only $\lesssim$0.3 mag over 50 years.  If FBS0107 
is indeed a symbiotic nova in a prolonged outburst, then this outburst is extremely slow 
and the photometric stability during outburst is very high.

\subsection{Other Possibilities}

The optical emission lines in FBS0107 resemble that of a high N$_e$ planetary nebulae (PN),
as is true for many symbiotics.  However, compared to the template PN spectra in Munari \& 
Zwitter (2002), the ratios Balmer /[OIII] and [OIII]/[OI] are each much larger in FBS0107
than in a PN.  

Guti\'{e}rrez-Moreno, Moreno \& Cortes (1995) provide a diagositic
diagram to separate PN from symbiotic stars, using the line ratios
R1 = [OIII]4363/H$\gamma$  vs R2 = [OIII]5007/H$\beta$.  The [OIII]
line strengths are ratioed to a nearby Balmer line partly to minimize
the effects of differential extinction, which varies slowly with
wavelength.  In our case it is also important to use such ratios
because they are nearly independent of the spectrophotometric calibration.
This unknown calibration, which is normally critical in
using emission line ratios for quantitative purposes, is also slowly varying
with wavelength.  Therefore we should still be able to apply R1 and R2 to FBS0107.
Note that it is not safe for us to use the ratio R3 = [OIII]5007/[OIII]4363
as described in Guti\'{e}rrez-Moreno et al. (1995) because of the larger wavelength
separation of the two features.

Guti\'{e}rrez-Moreno et al. (1995) find that in a plot of R1 vs R2
(their Figure~1) the region divides into 3 wedges emanating from the origin.  The
lower wedge (Region A) contains only PN, while the upper wedge (Region C) contains only
symbiotic stars.  The middle wedge (Region B) is lightly populated with young
PN.  From Table 3 we find  R1 = 0.06 and R2 = 0.36 for FBS-0107.   This point
falls very close to the origin of the diagnostic diagram, in a region where the demarcation
wedges are converging and there are very few objects.  This is because the Balmer lines 
in FBS0107 are very strong compared to [OIII], pushing the values of R1 and R2 to 
very small values.  It is the normalization process using nearby Balmer
lines (which we need to retain) that is causing the difficulty;  the ratio of the
two [OIII] lines is the important variable, and that ratio is within range of the plot.
Therefore we characterize the two dividing lines for the three wedges as 
R4 $\equiv$ R1/R2 = 0.17 for the upper line and R4 = 0.075 for the lower line.
For FBS0107, R4 = 0.17.  This places FBS0107 on the dividing line
between symbiotics and young PN, and quite some distance from the wedge containing
PN.

A major difficulty in associating FBS0107 with PN is that the photospheric features
in FBS0107 are not that of a PN central star (hot subdwarf), but rather of an early F giant.  
Furthermore we detect no nebulosity around FBS0107.

B[e] stars (Zickgraf 1998) are early-type emission line stars with low-excitation permitted 
and forbidden emission lines, plus the IR signature of hot dust.  They have similarities to 
both symbiotic stars and to compact PN.   As in symbiotics, the group is not
homogeneous and may represent more of a common phenomenon than a group having a
common physical nature.  Unlike symbiotics, they are not required to have a composite
spectrum, and may mostly be single stars.  The observational similarities between
B[e] stars and symbiotic binaries in outburst are strong because the 
M star in symbiotics is often hidden from view during eruption.  It seems likely that
there is cross-contamination between lists of symbiotics and B[e] stars, awaiting
clarification based on the binary nature of the object. 

B[e] stars are mostly B stars, so the late A/early F spectrum in FBS0107 is not a  good
fit to the B[e] grouping.  Another distinction may be between the ``hot dust'' in 
B[e] stars and the cooler dust in D type symbiotic stars.  The 2MASS colors of
FBS0107 are J-H = 0.27 and H-K$_S$ = 0.65.  Gummersbach et al. (1995) show
the location of B[e] stars in the two-color diagram J-H vs H-K.  (The difference
between H-K and H-K$_S$ (Carpenter 2001) is unimportant for our purposes.)
FBS0107 falls just outside the blue edge (in both colors) of the B[e] star grouping, 
which is an inconclusive result.

Finally, one other object in the Kopylov et al. (1988) list has similarities
to FBS0107.  This object (FBS0022-021 = Psc 3) has a blue continuum
and numerous emission lines, some being forbidden.  Zharikov et al.
(2004) suggested that this might be an $\eta$~Car-type object, partly
due to the Fe-rich nature of the spectrum.  However, unlike FBS0022-021, 
the Fe lines in FBS0107 are not particularly strong or numererous.

\section{Conclusions}

Although our conclusion is not as rugged as desired, our opinion is that
FBS0107 is most likely a symbiotic nova, seen in prolonged outburst.  This judgement is
partly due to the shortcomings of various competing identifications, but is mostly 
based on the presence of the early-F photosphere in the spectrum of FBS0107.  The
presence of this absorption spectrum is quite secure and is difficult to explain 
without invoking the outburst phase of a nova event.  The fact that FBS0107 deviates
somewhat from the properties of most symbiotic stars insofar as IR colors and emission line
ratios is tentatively attributed to the fact that FBS0107 is in prolonged
outburst, and therefore not representative of the majority of symbiotic binaries
which are found in quiesence.  
A somewhat worrisome implication of our interpretation is that FBS0107
must be an extreme member of the small class of symbiotic novae, having an outburst
time scale perhaps as long as that of AG Peg, and changing so slowly at present
that photometric variations are barely detectable.  

Symbiotic binaries have been considered as progenitors of Type Ia supernovae
(e.g. Munari \& Renzini 1982; Kenyon et al. 1993; Hachisu et al. 1999), especially the 
subclasses of recurrent novae (Mikolajewska 2008) and symbiotic novae, (e.g. Hernanz \& 
Jos\'{e} 2008).  Note however that there are considered  arguments against symbiotic 
binaries being a significant channel for SNIa (e.g. Hillebrandt \& Niemeyer 
2000; Iben \& Fujimoto 2008).  In any case, this aspect of symbiotic binaries may lend 
additional motivation to understanding the nature of FBS0107, especially if FBS0107 is
able to remain in very protracted thermonuclear outburst, as we are suggesting. 

Fortunately our symbiotic outburst hypothesis has fairly straightforward tests.  If 
FBS0107 is currently in symbiotic outburst, then photographic plate archives may provide 
a record of the pre-outburst
state.   Also, narrow-band imaging at high angular resolution may reveal nebulosity.
Non-optical observations can further discriminate among the possible candidates for
the nature of FBS0107; for example, it would be most interesting to know if FBS0107 is (or has
been) a supersoft x-ray source, indicative of steady nuclear burning.  Overall, it seems that 
one of our more definite conclusions is the all-too-common ending to observational papers 
that ``more observations are needed''.

\begin{deluxetable}{clcrrcc}
\tablenum{1}
\tablewidth{0pt}
\tablecolumns{7}
\tablecaption{Photometry Log}
\tablehead{\colhead{Set} & \colhead{UT Range} & \colhead{Tel} & \colhead{Exp time} 
& \colhead{\# Exps} & \colhead{Dur} &\colhead{Spacing}}
\startdata
1 & 2005-Nov-17    & Tenagra      & 95s  &  96 & 6.0 hr   & 140s  \\
  &                &              &      &     &          &       \\
2 & 2005-Nov-21    & Tenagra      & 95s  & 202 & 5.1 hr   & 140s  \\
  &                &              &      &     &          &       \\
3 & 2005-Nov-05 to & Tenagra      & 95s  &  37 & 45d      & 1-2d  \\
  & 2005-Dec-22    &              &      &     &          &       \\
  &                &              &      &     &          &       \\
4 & 2007-Aug-08 to & IN 1.25-m    & 240s & 122 & 514d     & 2-4d  \\
  & 2009-Jan-02    &              &      &     &          &       \\
   
\enddata
\end{deluxetable}
\clearpage

\begin{deluxetable}{ccccc}
\tablenum{2}
\tablewidth{0pt}
\tablecolumns{5}
\tablecaption{Spectroscopy Log}
\tablehead{\colhead{UT Date}  & \colhead{Tel} & \colhead{Range(\AA)}
& \colhead{Exp time(s)} & \colhead{\# Exps}}
\startdata
 2004-Aug-27 & ESO 3.6-m   & 3450-5240 & 600 & 5  \\
 2004-Aug-27 & ESO 3.6-m   & 4700-6700 & 600 & 18 \\    
 2005-Oct-25 & WIYN 3.5-m  & 5300-8200 & 600 & 5  \\
 2006-Jan-03 & KPNO 4-m    & 5600-7900 & 600 & 1  \\   
\enddata
\end{deluxetable}
\clearpage

\begin{deluxetable}{ccccc}
\tablenum{3}
\tablewidth{0pt}
\tablecolumns{5}
\tablecaption{Equivalent Widths of Lines in FBS 0107-082}
\tablehead{\colhead{Line}  & \colhead{ESO-blue} & \colhead{ESO-red}
& \colhead{WIYN} & \colhead{KPNO}}
\startdata
H8                     &  -4.8     &  \nodata  &  \nodata  &  \nodata  \\
CaII 3934              &  +0.9     &  \nodata  &  \nodata  &  \nodata  \\
H$\epsilon$            &  -3.9     &  \nodata  &  \nodata  &  \nodata  \\
HeII 4026              &  -1.4     &  \nodata  &  \nodata  &  \nodata  \\
H$\delta$              &  -8.0     &  \nodata  &  \nodata  &  \nodata  \\
H$\gamma$              &  -16.1    &  \nodata  &  \nodata  &  \nodata  \\
{[OIII]} 4363          &  -1.0     &  \nodata  &  \nodata  &  \nodata  \\
HeI 4388               &  -1.0     &  \nodata  &  \nodata  &  \nodata  \\
{[FeII]} 4414          &  -0.7     &  \nodata  &  \nodata  &  \nodata  \\
HeI 4471               &  -2.5     &  \nodata  &  \nodata  &  \nodata  \\
HeII 4686              &  -3.6     &  \nodata  &  \nodata  &  \nodata  \\
HeI 4713               &  -1.3     &  \nodata  &  \nodata  &  \nodata  \\
H$\beta$               &  -56.9    &  -60.0    &  \nodata  &  \nodata  \\
HeI 4922               &  -2.4     &  -3.4     &  \nodata  &  \nodata  \\
{[OIII]} 4959          &  -6.7     &  -7.4     &  \nodata  &  \nodata  \\
{[OIII]} 5007          &  -19.6    &  -21.7    &  \nodata  &  \nodata  \\
HeI 5015               &  -2.9     &  -4.2     &  \nodata  &  \nodata  \\
HeI 5876               &  \nodata  &  -9.9     &  -8.3     &  -8.6     \\
NaD1 5890              &  \nodata  &  +0.6     &  +0.3     &  +0.8     \\
NaD2 5896              &  \nodata  &  +0.5     &  +0.4     &  +0.4     \\
FeI 6164               &  \nodata  &  -1.9     &  -1.9     &  -2.3     \\
{[OI]} 6300            &  \nodata  &  -1.5     &  -1.2     &  -1.3     \\
H$\alpha$              &  \nodata  &  -453     &  -402     &  -424     \\
HeI 6678               &  \nodata  &  -5.4     &  -3.9     &  -1.8     \\
HeI 7065               &  \nodata  &  \nodata  &  -6.0     &  -6.6     \\
HeI 7281               &  \nodata  &  \nodata  &  -1.9     &  -2.0     \\
OI              7772   &  \nodata  &  \nodata  &  +1.2     &  +1.2     \\  
\enddata
\end{deluxetable}
\clearpage


\begin{figure}  
\epsscale{0.9}
\plotone{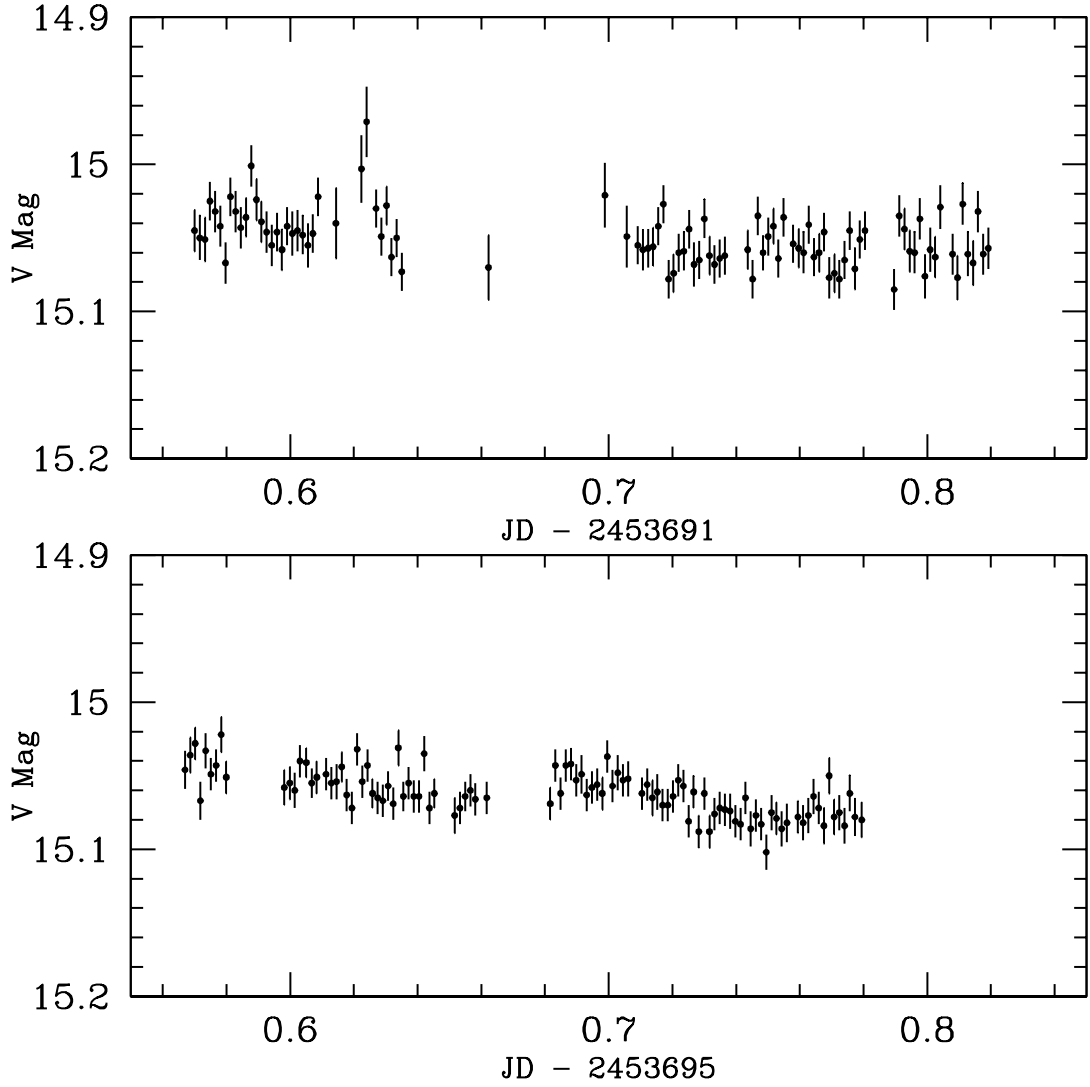}
\caption{Photometric sequences of FBS0107 on two nights in 2005.  These are
plots of Data sets 1 and 2 in Table 1.} 
\label{xx1}
\end{figure}
\clearpage

\begin{figure}  
\epsscale{0.9}
\plotone{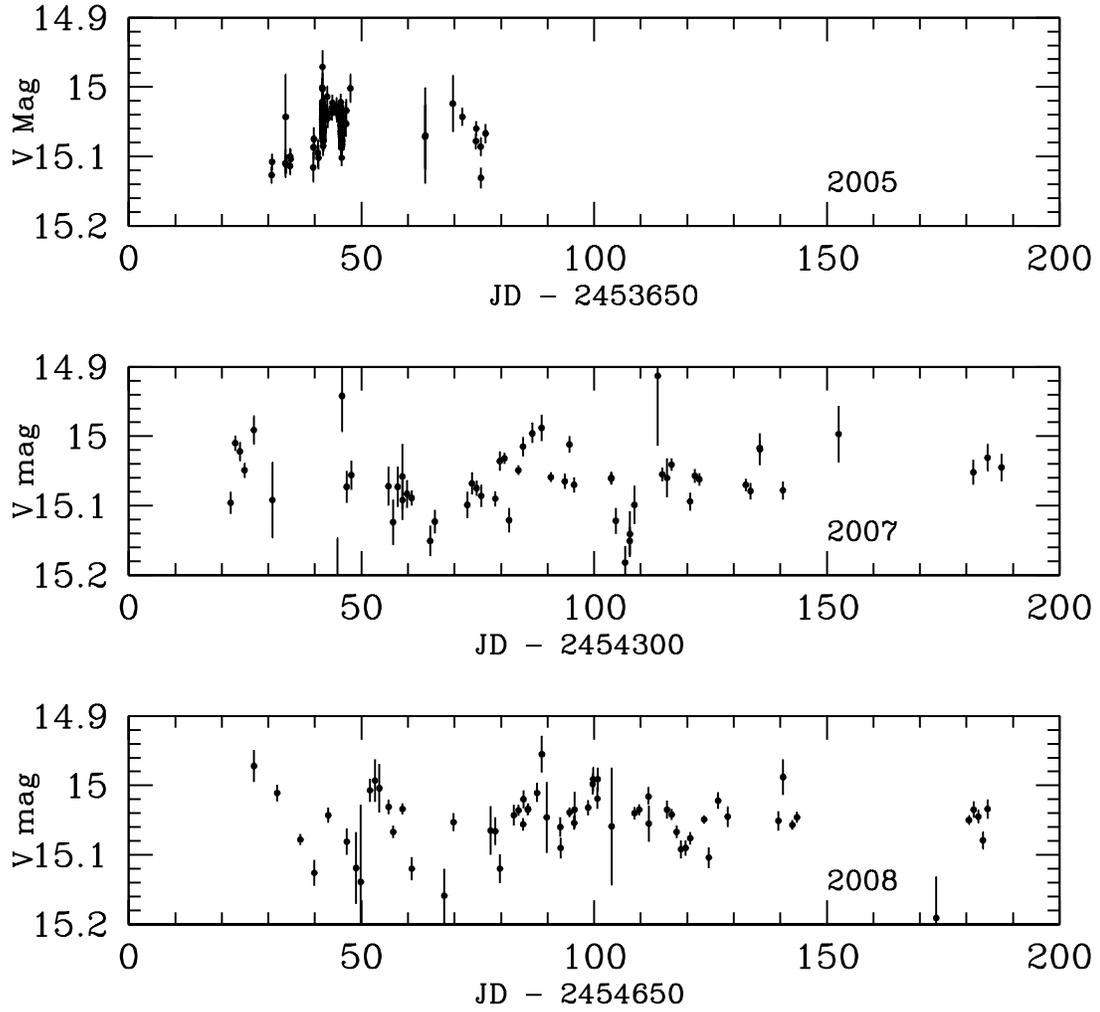}
\caption{Top panel: photometry of FBS-0107 over 1.5 months in 2005.  These
are plots of Data sets 1, 2, and 3 in Table 1.  Middle panel: photometry of 
FBS0107 over 1 season in 2007, which is part of Data set 4 in Table 1.
Lower panel: photometry of FBS0107 over 1 season in 2008, which is also part of
Data set 4 in Table 1.} 
\label{xx2}
\end{figure}
\clearpage

\begin{figure}  
\epsscale{0.9}
\plotone{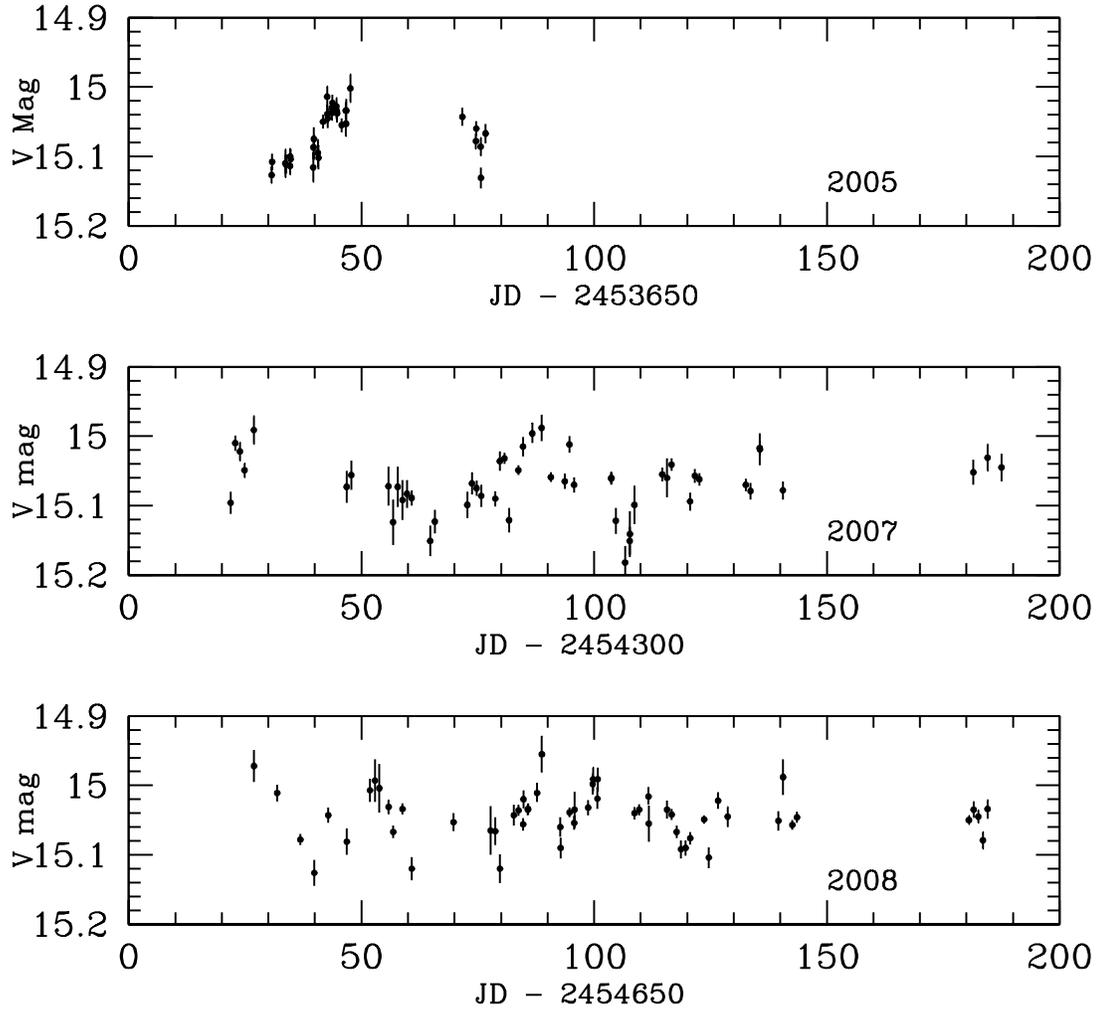}
\caption{Like Figure~2 except that points with errors exceeding 0.035 mag have been
deleted, and the single-night sequences (Data sets 1 and 2 in Table 1) have
been replaced by single mean points.  The
remaining points show more clearly a number of systematic changes in
brightness that occur over intervals of a few weeks.} 
\label{xx3}
\end{figure}
\clearpage

\begin{figure}  
\epsscale{0.85}
\plotone{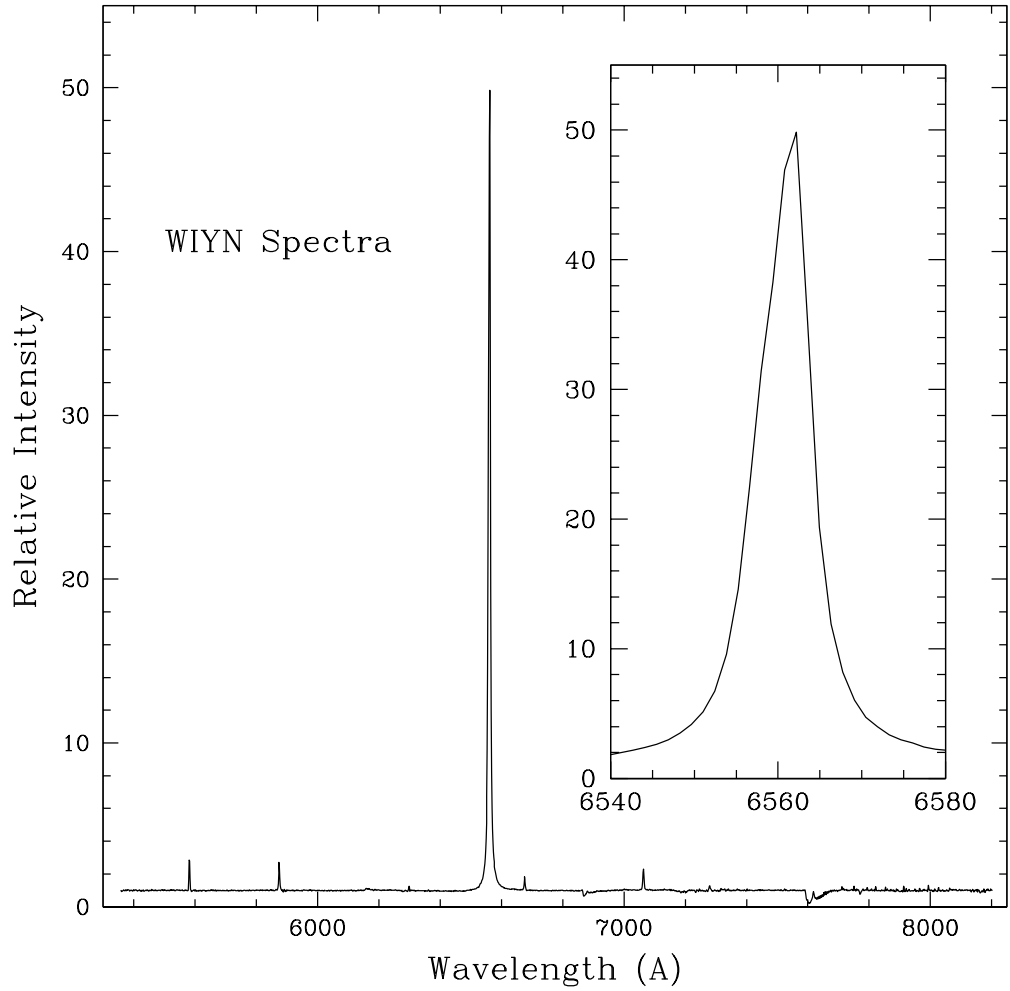}
\caption{Average of 5 WIYN exps of FBS~0107-082 on 2005-Oct-25 UT.  There has 
been no correction for telluric absorption features.  Also the correction
for sky emission lines is incomplete.  This has left some residual OH emission 
features redward of 7700\AA\ and some unknown contribution of night sky
[OI] 5577 to the HeI 5582 line.  The continuum is normalized to unity, and the 
inset shows H$\alpha$ on an expanded scale} 
\label{xx4}
\end{figure}
\clearpage

\begin{figure}  
\epsscale{0.85}
\plotone{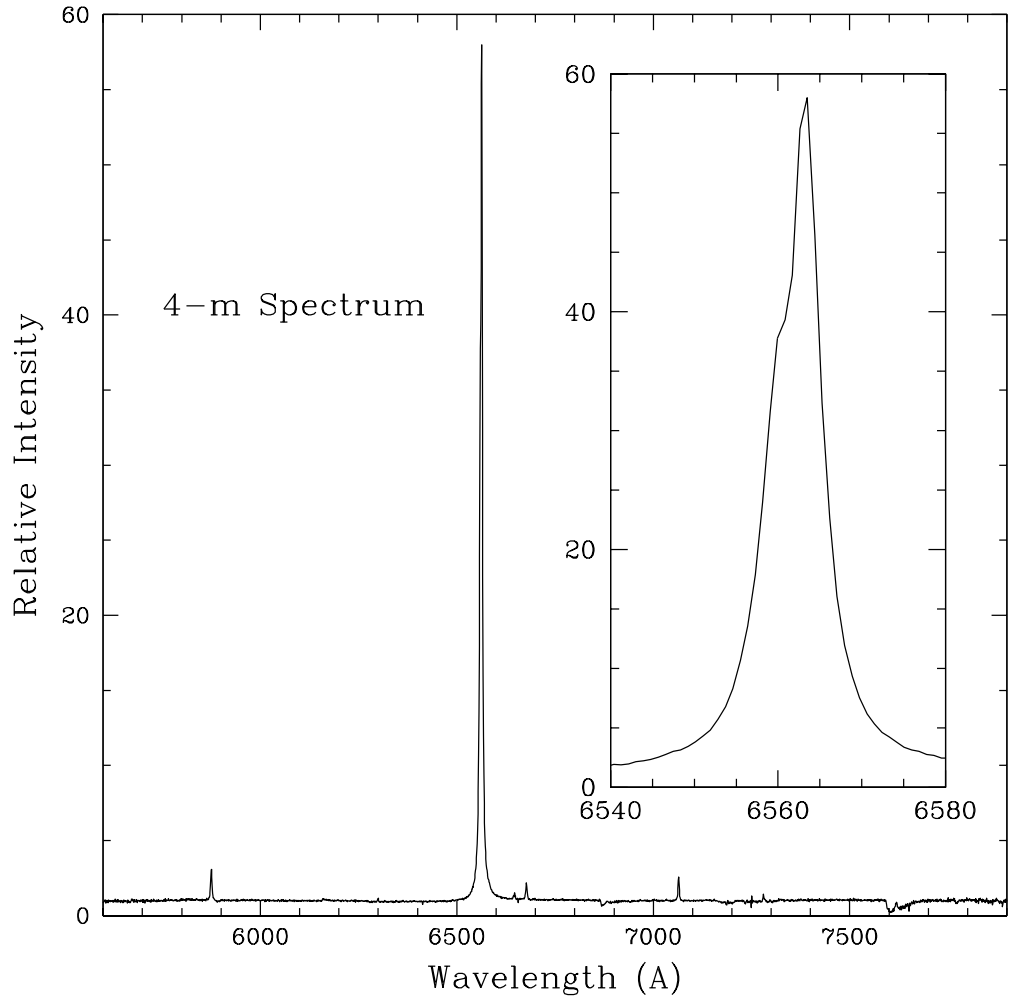}
\caption{Spectrum of FBS~0107-082 on 2006-Jan-03 UT using the KNPO 4-m.  There
has been no correction for telluric absorption features.  The continuum is
normalized to unity, and the inset shows H$\alpha$ on an expanded scale.}
\label{xx5}
\end{figure}
\clearpage

\begin{figure}  
\epsscale{0.9}
\plotone{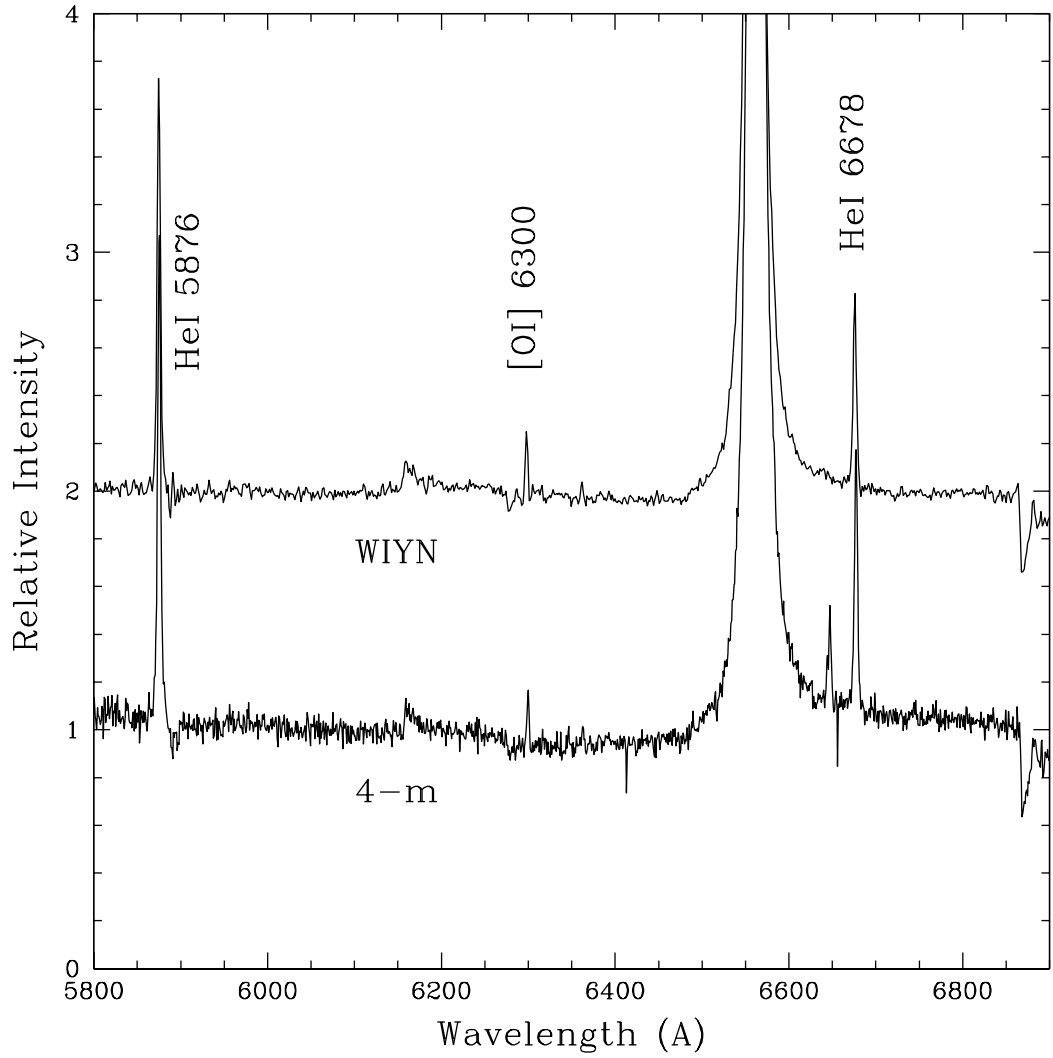}
\caption{The blue portions of the data in Figures~4 and 5, plotted for the
wavelengths in common and using an expanded intesity scale to emphasize 
the weaker features.  The apparent emission line near 6650\AA\ in the 4-m
spectrum is an artifact.}
\label{xx6}
\end{figure}
\clearpage

\begin{figure}  
\epsscale{0.9}
\plotone{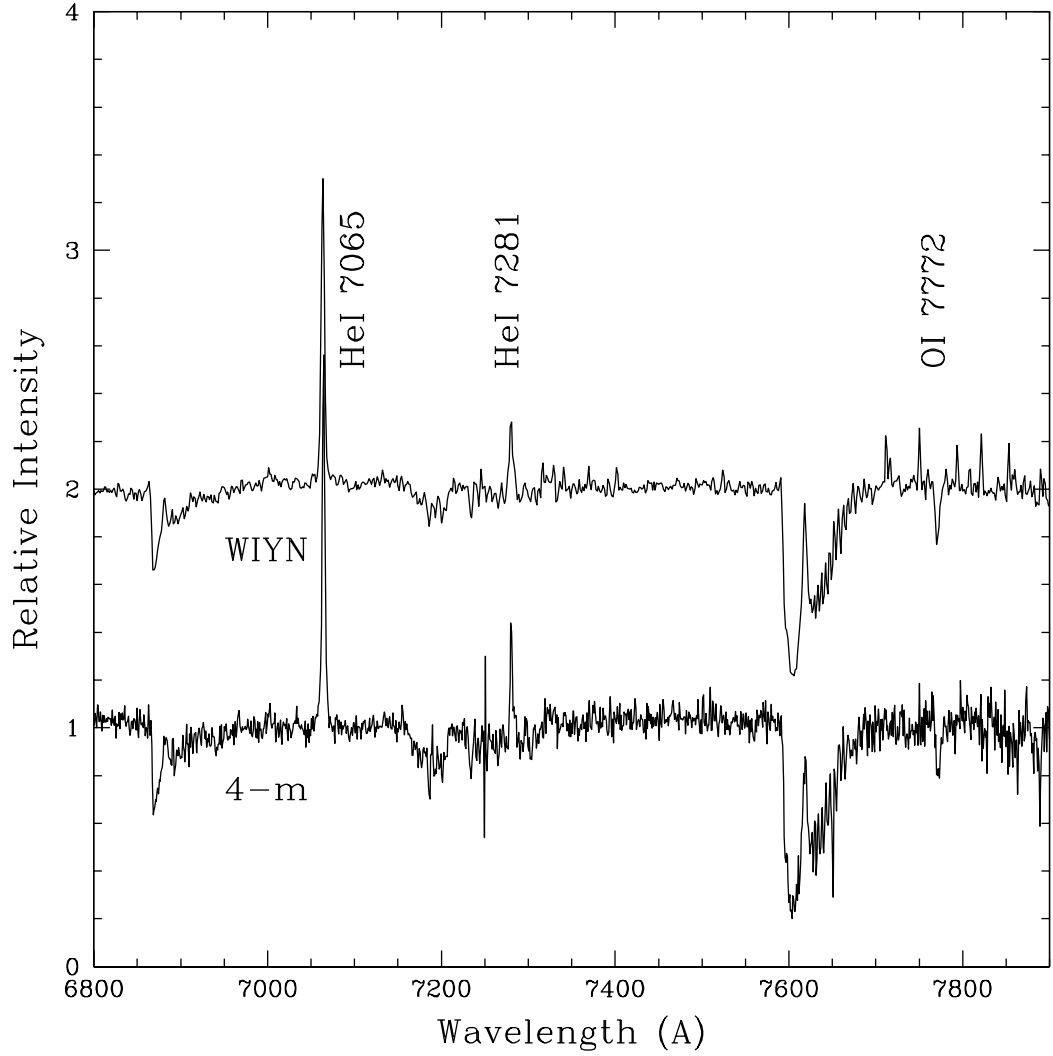}
\caption{The red portions of the data in Figures~4 and 5, plotted for the
wavelengths in common and using an expanded intensity scale to emphasize 
the weaker features.  Broad telluric absorption is seen near 7190 and
7620\AA.}
\label{xx7}
\end{figure}
\clearpage

\begin{figure}  
\epsscale{0.9}
\plotone{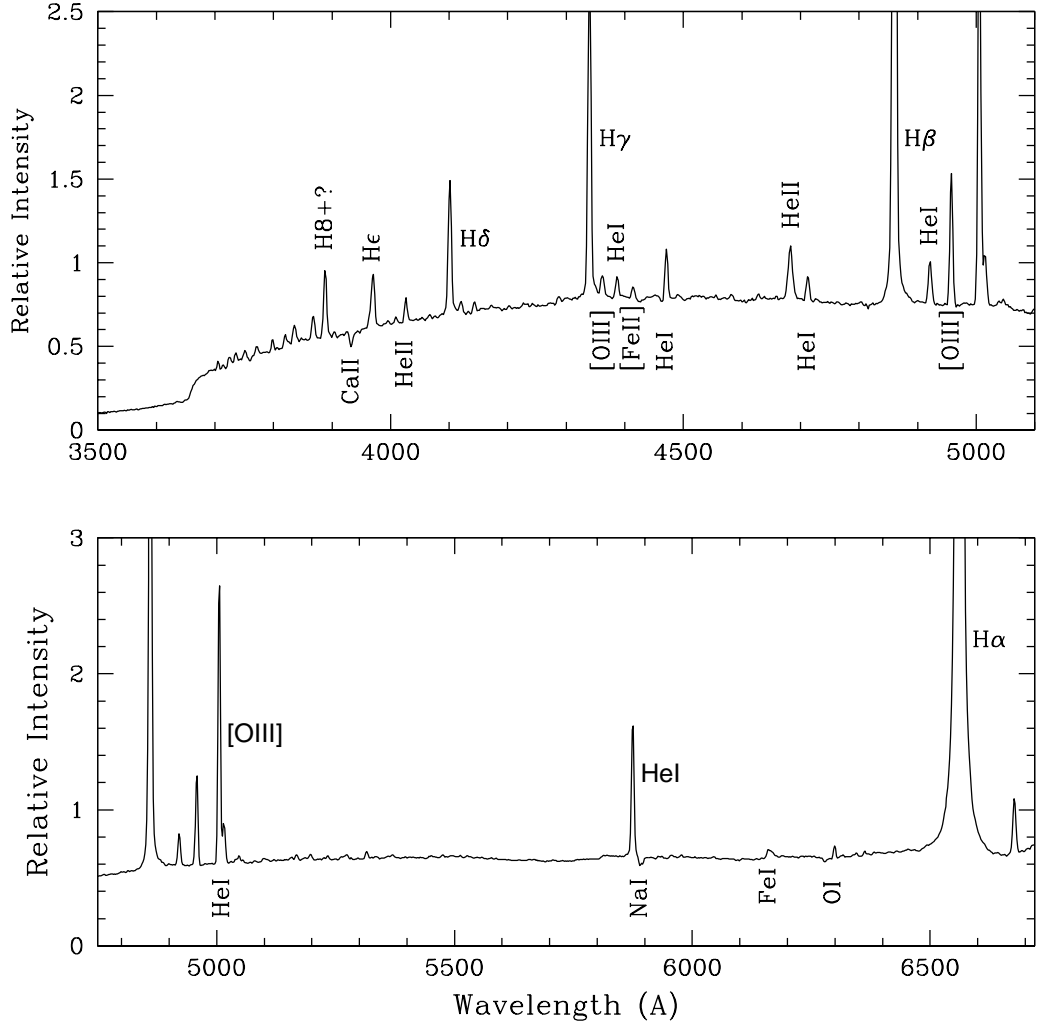}
\caption{The 2004 ESO spectra of FBS0107, dominated by Balmer emission,
He lines, and nebular lines of O and Fe.}
\label{xx8}
\end{figure}
\clearpage

\begin{figure}  
\epsscale{0.9}
\plotone{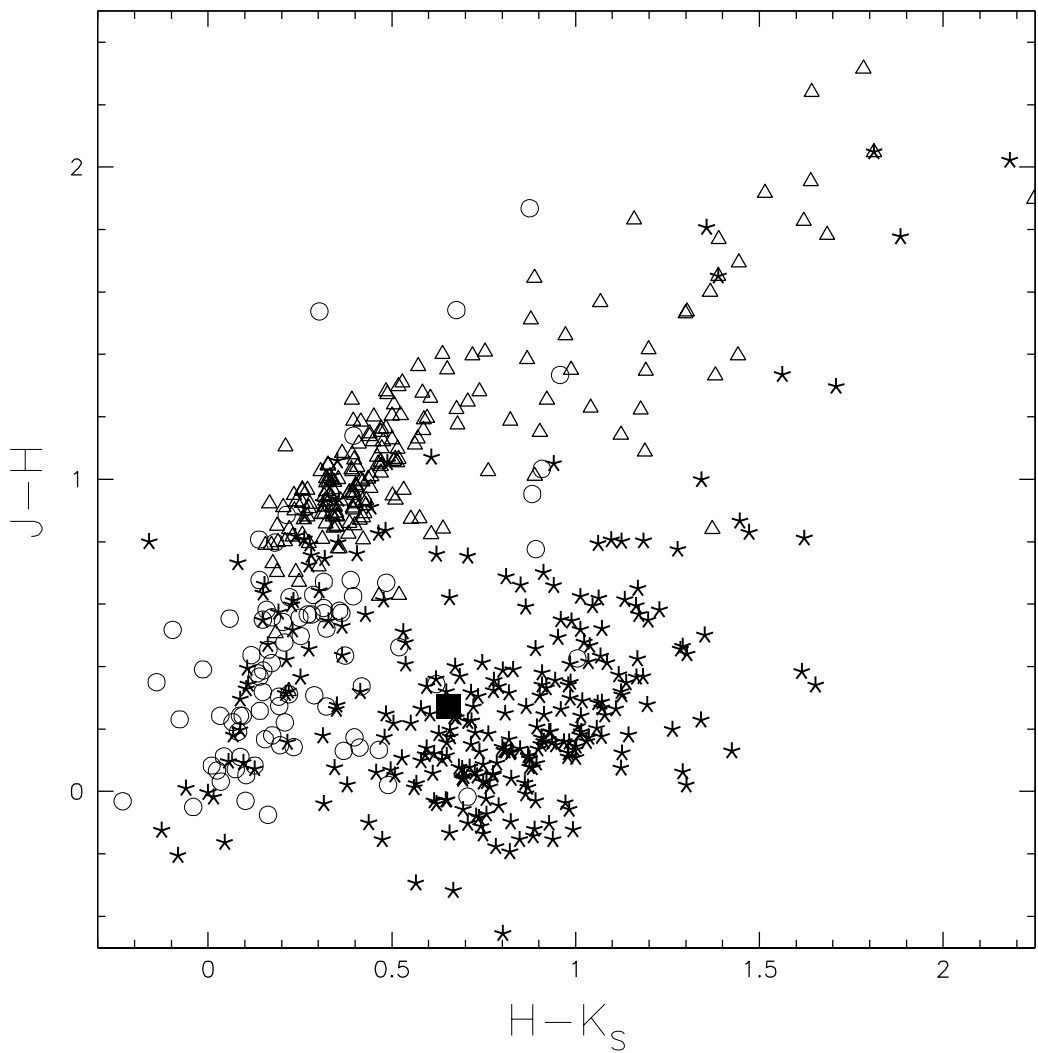}
\caption{2MASS color-color plot of CVs (open circles), planetary nebula (stars),
and symbiotic stars (open triangles).  The location of FBS0107 is indicated by
a large filled square.  Although FBS0107 has many of the optical spectral 
characteristics of symbiotic stars, its near-IR colors place in the region occupied 
by planetary nebulae.}
\label{xx10}
\end{figure}
\clearpage

\end{document}